# Leveraging Knowledge Graphs and Large Language Models to Track and Analyze Learning Trajectories

[1]Yu-Hxiang Chen, [1]Ju-Shen Huang, [1]Jia-Yu Hung, [1]Chia-Kai Chang *
[1]National Central University
ckchang@ncu.edu.tw

**ABSTRACT**: This study addresses the challenges of tracking and analyzing students' learning trajectories, particularly the issue of inadequate knowledge coverage in course assessments. Traditional assessment tools often fail to fully cover course content, leading to imprecise evaluations of student mastery. To tackle this problem, the study proposes a knowledge graph construction method based on large language models (LLMs), which transforms learning materials into structured data and generates personalized learning trajectory graphs by analyzing students' test data. Experimental results demonstrate that the model effectively alerts teachers to potential biases in their exam questions and tracks individual student progress. This system not only enhances the accuracy of learning assessments but also helps teachers provide timely guidance to students who are falling behind, thereby improving overall teaching strategies.

## 1    INTRODUCTION

Tracking and analyzing students' learning trajectories has become crucial in contemporary education (Ellis et al., 2014). Educational service platforms have already been implemented in industry, and academic research focuses on developing tools to explain and observe learning behaviors. For example, in 2012, Anna Lea Dyckhoff, Dennis Zielke, and others proposed the Exploratory Learning Analytics Toolkit (eLAT), which provides teachers with a user-friendly interface to explore students' learning activities and assessment results through data visualization, allowing them to reflect on and improve teaching strategies.

Similarly, José Michel Fogaça Vieira et al. proposed various methods of representing learning trajectories. However, these approaches are often limited to data display and are not widely applicable across different academic subjects. Therefore, we devised a strategy based on knowledge graph analysis that enables teachers to grasp students' learning progress better. For instance, it can monitor the extent of students' curriculum coverage and observe changes over time, providing insights into their learning trajectories.

This study introduces a system that leverages knowledge graphs built from large language models (LLMs) to analyze learning materials and track students' progress. By transforming the materials into a structured list of nodes and relationships, individualized knowledge graphs are generated for each student, integrating exam data to assess academic performance and teaching effectiveness. Applied to an introductory Python programming course at a national university in Taiwan, the system identified gaps in exam coverage and student progress, helping teachers adjust the scope of exams and providing targeted support to students lagging.

## 2    METHOD

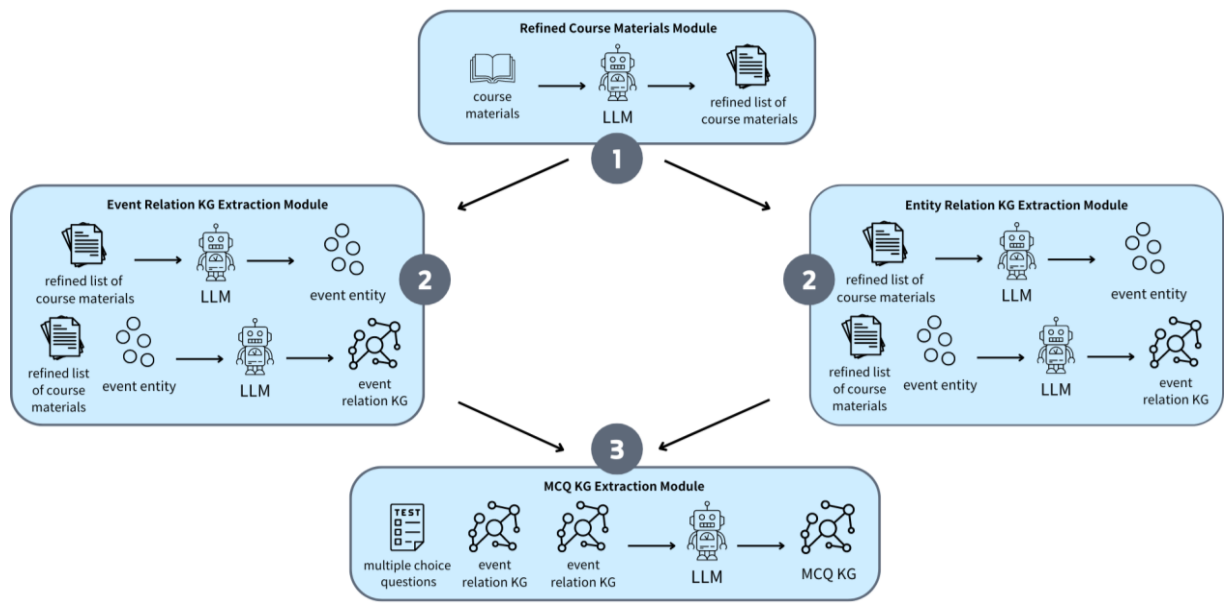

**Figure 1: Multiple-choice Questions Knowledge Graph Construction Framework Diagram**

As shown in the first step of Figure 1, we converted the course materials into text files. We used a Large Language Model (LLM) to create a refined list by removing unnecessary conjunctions and particles. This structured the content into concrete and meaningful data, making it more organized and interpretable for the subsequent construction and analysis of the knowledge graph, thus improving the accuracy and effectiveness of learning trajectory measurement.

In the second step of Figure 1, following Ling Feng Zhong et al., we extracted nodes (entities) from the refined list of materials, categorizing them as general nodes (people, objects, time, places) and event nodes (specific events). Using Shuang Yu et al.'s method, we utilized the LLM to automatically generate relationships (entity relations) between nodes, enhancing efficiency and accuracy. We then input both the general nodes and refined list into the LLM to create relationships, mainly verbs or prepositions connecting two nodes. Event nodes were also processed similarly to establish causal and sequential connections, forming a complete knowledge graph.

In the third step of Figure 1, we tracked students' learning trajectories using quizzes, midterms, and final exams to build their knowledge paths. Multiple-choice questions and answers from these assessments were input into the LLM and knowledge graph. The Chain-of-Thought (COT) process in Prompt Engineering enabled the LLM to match these questions to the corresponding edges in the knowledge graph, allowing a detailed mapping of student learning progress.

Once we had determined which edges in the knowledge graph corresponded to each question, we could create a personalized knowledge graph for each student to record their learning trajectory. When students correctly answered a question, we marked the corresponding edge in their knowledge graph. Through this complete learning trajectory-building process, we could study the changes in students' knowledge paths and use them to evaluate their abilities and the effectiveness of the course. Below are four aspects that can be explored in research:

- Changes in the knowledge graph correspond to different score groups.
- Identifying key knowledge points to determine which gaps lead to difficulty answering specific questions, thus causing learning bottlenecks.
- Changes in the coverage of knowledge points across the class are needed to assess whether students have mastered all the knowledge covered by the course after the teacher's instruction.
- Evaluating whether the test comprehensively assesses the knowledge students learn in the classroom.

## 3  CASE STUDY

### 3.1  Knowledge Node Coverage Warning and Cognitive Bias System for Instructors

Our framework model was implemented in an experimental research study on an introductory Python Programming course at a national university in Taiwan. A total of 47 students participated fully in the study. During the course, each student completed three standardized and unbiased multiple-choice assessments designed by the course instructor based on the curriculum and related to fundamental Python programming skills. In Figure 2, the color differences reflect knowledge point coverage across testing phases. Green dots in the Pre Test represent foundational knowledge assessed before instruction, while purple and blue dots in the Midterm and Post Test indicate knowledge introduced or reinforced during teaching. This highlights curriculum progression and helps identify gaps or newly emphasized concepts.

The study results are shown in Figure 3, which highlights part of the knowledge graph depicting the distribution of knowledge nodes across the three assessments. The percentage of knowledge nodes covered was 6.1% in the pre-test, 8.8% in the mid-term exam, and 6.1% in the post-test. It was noted that the knowledge nodes in the pre-test and post-test overlapped significantly with those in the mid-term exam. This suggests that the instructor's selection of knowledge points may have been influenced by selective attention, a phenomenon where focus is unintentionally directed toward specific areas, potentially overlooking other important knowledge nodes. As a result, the three assessments did not adequately cover the entire scope of the course content. Our system allows for early detection of such gaps, helping instructors adjust the scope and content of future assessments.

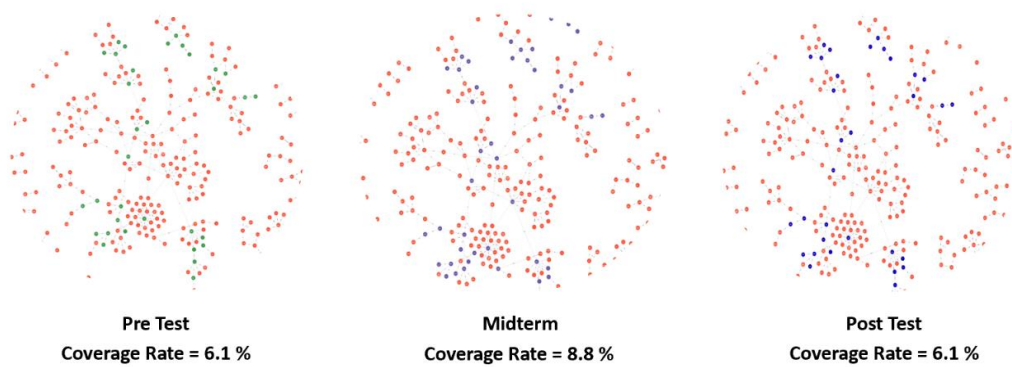

**Figure 2: Intersection of Knowledge Graph Coverage for Three Assessments and the Overall Python Course.**

## 3.2   Student Performance Warning System

We also analyzed the growth rate of knowledge nodes for individual students across the three assessments and compared it to the class average. Figure 3 compares a student's knowledge graph coverage with the class average. The student's coverage rate for knowledge nodes in the mid-term exam was 79.4%. In the figure, the red areas represent the course's knowledge points, the yellow areas indicate the knowledge points already mastered by the student, and the green areas indicate the knowledge points the student lags the class average in mastering. Using our system, students can identify areas requiring improvement and receive targeted alerts based on their level of knowledge deficiency.

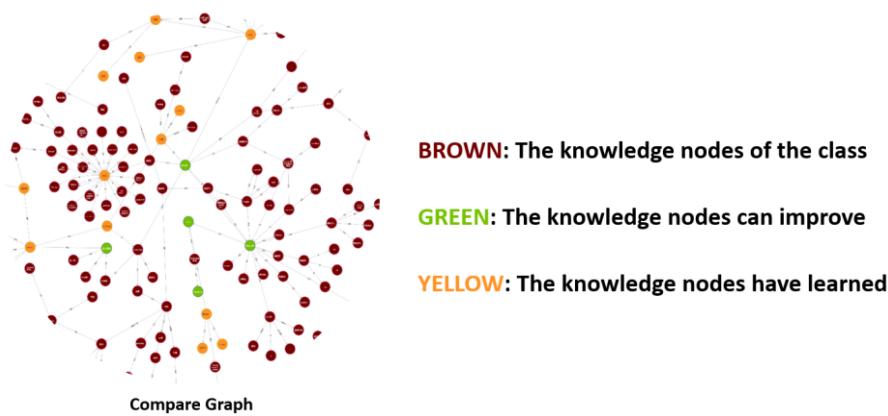

**Figure 3: Knowledge Graph Comparison Between an Individual Student's Mastery and the Class Average.**

## 4   CONCLUSION

This paper proposes a knowledge graph-based method using LLM to track and analyze students' learning trajectories, addressing the issue of incomplete coverage in traditional assessments. Individualized knowledge graphs are generated by converting teaching materials into structured data and integrating them with students' test results, mapping learning progress, and identifying gaps. The system helps educators adjust exam content, track performance, and support students, improving assessment accuracy and teaching strategies.